\documentclass[a4j,leqno]{article}
\newtheorem{thm}{Theorem}[section]
\newtheorem{prop}[thm]{Proposition}

\newtheorem{pf}{Proof}

\newtheorem{exmp}{Example}[section]
\newtheorem{defn}[thm]{Definition}
\usepackage{amssymb}
\usepackage{latexsym}
\usepackage{enumerate}
\usepackage{amsmath}
\usepackage{amsfonts}
\usepackage[mathscr]{eucal}
\usepackage{amscd}
\usepackage{indentfirst}
\usepackage{txfonts}
\usepackage{ulem} 
\numberwithin{equation}{section}

\usepackage{amssymb}
\usepackage{color}

\begin{document}
\begin{center}
{\Large\bf Formal Deformation Quantization for Super Poisson Structures on 
Super Calabi-Yau Twistor Spaces}
\end{center}
\begin{center}
Dedicated to Professor Yoshiaki Maeda on his 70th birthday 
\end{center}
\begin{center}
{\large Tadashi TANIGUCHI}\\
Gunma National College of Technology, Maebashi, 371-8530, JAPAN \\ 
{\large Naoya MIYAZAKI}\\
Keio University, Yokohama, 223-8521, JAPAN\\ 
{\large Yuji HIROTA}\\ 
Azabu University, Sagamihara, 252-5201, JAPAN\\
\end{center}


{\abstract 
We endow super Poisson structures 
with super Calabi-Yau manifolds 
by using super twistor double fibrations．
Moreover 
we define the structure of deformation quantization for such 
super Poisson manifolds. 
}

\section{Introduction}

It has been shown that \cite{aganagic-vafa} there exists the relationship 
between ${\Bbb L}^{5\,|\, 6}$ and ${\Bbb P}^{3\,|\, 4}$ analogous to Mirror symmetry in some sense.
It is known that the Calabi-Yau condition is 
the most important in order to study of Mirror symmetry.  
In the present paper, we are mainly concerned with super Calabi-Yau twistor spaces. 
It is known that Wolf \cite{wolf} introduced super Calabi-Yau twistor spaces. 

As mentioned in abstract, we introduce super Poisson structures on 
super Calabi-Yau twistor spaces. Then, we also study deformation quantization of 
super Calabi-Yau twistor spaces via the super Poisson structures. 
When $N=0$, super Poisson manifolds are reduced to 
the classical Poisson manifolds．

The first purpose of the paper is to demonstrate that the 
super twistor double fibration plays a crucial role in our construction. 
We construct even super Poisson structures on super Calabi-Yau twistor spaces as in the case of 
classical dimension $= 3$. 

The second one is to study deformation quantization of them.
We show that even super Calabi-Yau twistor Poisson spaces can be deformation quantizable.
In the topics, we shall exhibit Propositions \ref{boson-fermion-commutation-relation1}, 
\ref{boson-fermion-commutation-relation2} 
and Theorem \ref{boson-fermion-commutation-relation3}. 
It is also able to construct {\bf odd} super Calabi-Yau twistor Poisson spaces in the way similar to 
and seems that those spaces can be deformation quantizable. 

To end this section, we mention deformation quantization 
in the smooth and algebraic categories
(cf. \cite{bffls, dl, fedosov, k, omy, 
ommy1, ommy2, ommy3, miyazaki, yoshioka}).  
Kontsevich showed in his celebrated paper \cite{k} that 
any Poisson manifold can be deformation quantizable in smooth category with the formal parameter
, while Miyazaki \cite{miyazaki} discussed deformation quantizations 
in algebraic/analytic categories on the basis of the assumption that 
\[\begin{array}{ll}
&\overleftarrow{\partial}_{i_1}\omega^{i_1j_1}\overrightarrow{\partial}_{j_1}
\overleftarrow{\partial}_{i_2}\omega^{i_2j_2}\overrightarrow{\partial}_{j_2}
\cdots
\overleftarrow{\partial}_{i_\ell}\omega^{i_\ell j_\ell}\overrightarrow{\partial}_{j_\ell}
\\
=&\overleftarrow{\partial}_{i_1}\overleftarrow{\partial}_{i_2}\cdots
\overleftarrow{\partial}_{i_\ell}
\omega^{i_1j_1}\omega^{i_2j_2}\cdots \omega^{i_\ell j_\ell}
\overrightarrow{\partial}_{j_1}\overrightarrow{\partial}_{j_2}
\cdots \overrightarrow{\partial}_{j_\ell}
\end{array}
\]
in his first trial. The convergence of star product and the star exponential were discussed in 
\cite{ommy1, ommy2, ommy3}.
\vspace{0.5cm}

The paper consists of five sections. 
In Section 2, we give a short overview of theory of supermanifold, 
and mention fundamental examples which we need in this paper. 
In Section 3, we consider formal deformation quantization for 
super Poisson twistor spaces. 
In Section 4, we study other examples. 
Finally, in Section 5, we study glueing problem of resulting algebras 
via formal deformation quantization.

\section{Preliminaries}

A theory of supermanifold is based on super (or $\mathbb{Z}_2$-graded) algebra. In the section, we review quickly a supermanifold which we will use 
throughout the paper, supposing the reader has fundamental knowledge concerning superalgebras. 
For the further information for superalgebras and supermanifolds, we refer the reader to 
\cite{lebr, mani, manin, taniguchi1, taniguchi, ward-wells}. 
\vspace{0.3cm}

Let us denote by $\mathcal{C}_M^\infty$ a sheaf of smooth functions on a smooth manifold $M$. We remark that 
every smooth manifold $M$ is described as a ringed space $(M,\,\mathcal{C}^\infty_M)$. A supermanifold is defined in terms of a ringed space 
in the same way as the ordinary one. 

\begin{defn}\label{sec2:defn_supermanifold}
A supermanifold of dimension $(n \,|\, N)$ is a ringed space $(M,\,\mathcal{A}_M)$, where $M$ is a smooth manifold of dimension $n$, and 
$\mathcal{A}_M=(\mathcal{A}_M)_0\oplus (\mathcal{A}_M)_1$ is a sheaf of supercommutative rings on $M$, called the structure sheaf on $M$, 
which satisfies the following conditions {\rm :} 
\begin{enumerate}[\quad \rm (1)]  
\item For a nilpotent ideal sheaf $\mathcal{N}_M\coloneqq(\mathcal{A}_M)_1 +{(\mathcal{A}_M)_1}^2$ of $\mathcal{A}_M$, $(M,\,\mathcal{A}_M /\mathcal{N}_M)$ 
is a smooth manifold. 
\item ${\cal N}_M/{{\cal N}_M}^2$ is a locally free $\mathcal{C}_M^\infty$-module of rank $N$.  
\item $\mathcal{A}_M$ is locally isomorphic, as a sheaf of supercommutative algebras, to the exterior algebra sheaf  $\bigwedge^{\bullet}(\mathcal{N}_M/{{\cal N}_M}^2)$. 
\end{enumerate}
\end{defn} 

We sometimes use the notation $\hat{M}$ for a supermanifold $(M,\,\mathcal{A}_M)$ throughout the paper. 
For a supermanifold $\hat{M}$, a local section $f$ of the structure sheaf is written in the 
following way: 
\begin{equation}\label{local-represent}
f=\sum_{k=1}^{N} 
\sum_{1 \leqq i_1 < i_2 < \cdots < i_k \leqq N} 
f_{i_1 i_2 \ldots i_k} (z)\, \theta^{i_1} \theta^{i_2} \cdots \theta^{i_k} , 
\end{equation}
where $f_{i_1 i_2 \cdots i_k}(z)$ 
are smooth functions on $M$ with local coordinates $z_1,z_2, \cdots ,z_n$ and where 
$\theta^1, \theta^2 , \cdots ,\theta^N $ are odd variables so that they 
are local generators of ${\cal N}_M/{{\cal N}_M}^2$. 
We use the notation 
$$(z\,|\, \theta)=(z_1, z_2, \cdots ,z_n \,|\, \theta^1, \theta^2, \cdots, \theta^N)$$  
for a local coordinate of a supermanifold $\hat{M}$. 

Analogously, one can define the notion of a complex supermanifold by taking a complex manifold as an underlying smooth manifold and replacing $\mathcal{C}_M^\infty$ 
with a sheaf of holomorphic functions in Definition \ref{sec2:defn_supermanifold}. 
Noting the fact, let us recall the definition of super twistor manifold (see \cite{miyazaki}).
\begin{defn}
$(3\,|\, N)$-dimensional complex super manifold $\hat{Z}$ is called 
a {\rm super twistor space} if the following conditions are satisfied.  
\begin{enumerate}[\quad\rm (1)] 
\item $p: Z\longrightarrow {\Bbb P}^1$ is a holomorphic fiber bundle.
\item $Z$ has a family of holomorphic section of $p$ 
whose normal bundle ${\cal N}$ is isomorphic to 
${\cal N}\cong {\cal O}_{{\Bbb P}^1}(m_1)\oplus {\cal O}_{{\Bbb P}^1}(m_2)\oplus \displaystyle\bigoplus_{i=1}^{N} \Pi {\cal O}_{{\Bbb P}^1}(n_i)$, 
where $\Pi$ denotes the parity change functor.  

\end{enumerate}
\end{defn}


\par\medskip\noindent
\begin{exmp}

\noindent
$\rm{(I)}$\,
The typical example is the real 
(or complex) super vector space ${\Bbb R}^{n\,|\, N}$
(or ${\Bbb C}^{n\,|\, N}$)  
which can be defined by 
 \[ {\Bbb R}^{n\,|\, N}=\Bigl( {\Bbb R}^n, \,
 \bigwedge^{\bullet} ({\Bbb R}^N \otimes_{\Bbb R} {\cal O}_{{\Bbb R}^n })\Bigr),\]
\[ {\Bbb C}^{n\vert N} = \Bigl({\Bbb C}^n, \,
\bigwedge^{\bullet} ({\Bbb C}^N \otimes_{\Bbb C} {\cal O}_{{\Bbb C}^n })\Bigr). \]
It is easy to see that ${\Bbb R}^{n\,|\, N}$ is isomorphic to
$\Bigl({\Bbb R}^n, {\cal O}(S^{\cdot}({\Bbb R}^n) 
\otimes\bigwedge^{\bullet}{\Bbb R}^N)\Bigr)$. \\

\noindent
$\rm{(II)}$\, 
A complex super projective space of dimension $(n\,|\, N)$ 
is defined by  \[ {\Bbb P}^{n\,|\, N}=\Bigl({\Bbb P}^n,\, \bigwedge^{\bullet}({\Bbb C}^N 
\otimes_{\Bbb C} {\cal O}_{{\Bbb P}^n}(-1))\Bigr).\]
We denote by $ {\cal O}_{{\Bbb P}^{n\,|\, N}}$ the structure sheaf  
$\bigwedge^{\cdot} ({\Bbb C}^N \otimes_{\Bbb C}{\cal O}_{{\Bbb P}^n}(-1))$
of ${\Bbb P}^{n\,|\, N}$. 
A local section of ${\cal O}_{{\Bbb P}^{n\,|\, N}}$ is represented 
as in {\rm (\ref{local-represent})} via a degree $(-k)$ homogeneous element 
$f_{i_1 i_2\cdots i_k}(z)$. 
For example, if $(n\,|\,N)=(3\,|\,2)$, we see that 
${\cal O}_{{\Bbb P}^{3\,|\, 2}}$ is decomposed in the following way: 
\[ 
{\cal O}_{{\Bbb P}^{3\,|\, 2}}\cong {\cal O}_{{\Bbb P}^3} \oplus 
\Pi {\cal O}_{{\Bbb P}^3}(-1)\oplus \Pi {\cal O}_{{\Bbb P}^3}(-1)\oplus 
{\cal O}_{{\Bbb P}^3}(-2),
\]
where $\Pi$ denotes the parity change functor as mentioned before.
We define the line sheaf of degree $d$, denoted by ${\cal O}_{{\Bbb P}^{n\,|\, N}} (d)$ as 
${\cal O}_{{\Bbb P}^{n\,|\, N}} (d) = 
{\cal O}_{{\Bbb P}^n}(d)\otimes {\cal O}_{{\Bbb P}^{n\,|\, N}}$, 
where ${\cal O}_{{\Bbb P}^n}(d)$ stands for 
the sheaf of germs of locally defined, holomorphic and homogeneous degree $d$. 

The first Chern class is given by
\[ c_1 ({\Bbb P}^{n \,|\, N})=
c_1 ({\cal O}_{{\Bbb P}^{n \,|\, N}} \otimes {\Bbb C}^{n+1}) - 
c_1 ({\cal O}_{{\Bbb P}^{n \,|\, N}}\otimes {\Bbb C}^N) = (n+1-N)x, \]
where $x=c_1 ({\cal O}_{{\Bbb P}^{n\,|\, N}}(1))$.
Hence a super twistor space 
${\Bbb P}^{n\,|\, N}$ 
{\bf\rm might admit the structure of a Calabi-Yau supermanifold (or might satisfy a Calabi-Yau condition) if and only if $N=4$}.

\par\bigskip
\noindent
$\rm{(III)}$\,
A weighted super projective space \cite[p.40]{wolf} 
is defined by
\[
{\Bbb W}{\Bbb P}^{n\,|\, N} [k_1,k_2,\cdots , k_{n+1}\, |\, l_1,l_2,\cdots ,l_N ]
=  ({\Bbb W} {\Bbb P}^n [k_1,k_2,\cdots ,k_{n+1}] , 
{\cal O}_{{\Bbb W}{\Bbb P}^{n\,|\, N}} ), 
\]
where 
\[  {\cal O}_{{\Bbb W}{\Bbb P}^{n\,|\, N}} 
= \bigwedge^{\cdot} ({\cal O}_{{\Bbb W}{\Bbb P}^n} (-l_1)\oplus \cdots \oplus {\cal O}_{{\Bbb W}{\Bbb P}^n} (-l_N )) .\]
Note that with the definitions above, we have   
\[ {\Bbb W}{\Bbb P}^{n\,|\, N} 
[\overbrace{1,1, \cdots , 1}^{n+1} \vert \overbrace{1,1,\cdots ,1}^N ] = {\Bbb P}^{n\,|\, N} .\]
The first Chern class is given by
\[ c_1 ({\Bbb W}{\Bbb P}^{n\,|\, N})=(\sum_{i=1}^{n+1}k_i - \sum_{j=1}^{N} l_j )x, \]
$x=c_1 ({\cal O}_{{\Bbb W}{\Bbb P}^{n\,|\, N}}(1))$.
Hence, for appropriate numbers $k_i$ and $l_j$, the weighted projective super space ${\Bbb W}{\Bbb P}^{n\,|\, N}$
becomes a Calabi-Yau supermanifold.\\

Let us now consider an open subset of ${\Bbb W}{\Bbb P}^{3\,|\, N} [k_1,k_2,1,1\,|\, l_1,l_2,\cdots ,l_N ]$ defined by
\begin{equation}
\label{open-wp} 
\begin{array}{ll} 
& {\cal WP}^{3|N} [k_1,k_2,1,1 \,|\, l_1,l_2,\cdots ,l_N] \\ 
= & {\Bbb W}{\Bbb P}^{3\,|\, N} [k_1,k_2, 1, 1 \,|\, l_1,l_2,\cdots , l_N ] 
\setminus 
{\Bbb W}{\Bbb P}^{1\,|\, N} [k_1,k_2,1, 1 \,|\, l_1,l_2,\cdots , l_N ]. 
\end{array}
\end{equation}
This can be identified with the holomorphic fiber bundle
\[ 
{\cal O}_{{\Bbb P}^1}(k_1) \oplus {\cal O}_{{\Bbb P}^1}(k_2) \oplus \bigoplus_{j=1}^N \Pi {\cal O}_{{\Bbb P}^1}(l_j) 
\longrightarrow {\Bbb P}^1, 
\]
and as such it can be covered by two patches (cf. \cite{wolf}).

When $k_1=k_2=l_1=l_2=\cdots =l_N=1$,
\[ 
{\cal P}^{3\,|\,N}= {\cal O}_{{\Bbb P}^1}(1) \oplus {\cal O}_{{\Bbb P}^1}(1) \oplus \bigoplus_{j=1}^N \Pi {\cal O}_{{\Bbb P}^1}(1), 
\quad 
{\cal P}^{3\,|\,N} \cong {\Bbb P}^{3\,|\,N}\setminus {\Bbb P}^{1\,|\,N}. 
\] 

In this case, Calabi-Yau condition is of the $k_1+k_2-(l_1+\cdots l_N)+2=0$.\\
In particular we consider the case of $N=2$, that is,  

\[{\cal WP}^{3\,|\,2} [p,q]={\Bbb W}{\Bbb P}^{3\,|\, 2} [1, 1, 1, 1 \,|\, p,q ]\setminus 
{\Bbb W}{\Bbb P}^{1\,|\, 2} [1, 1 \,|\, p,q ]. \]
This space can be identified with  the holomorphic fiber bundle
 \[  {\cal O}_{{\Bbb P}^1}(1) \oplus {\cal O}_{{\Bbb P}^1}(1) \oplus 
\Pi {\cal O}_{{\Bbb P}^1}(p)\oplus \Pi {\cal O}_{{\Bbb P}^1}(q)\longrightarrow {\Bbb P}^1. \]
For the particular combination $p+q=4$, 
it is a Calabi-Yau supermanifold.\\
There are three types of such Calabi-Yau supermanifolds as follows:
\[ {\cal WP}^3\,|\,2 [1,3],\qquad  
{\cal WP}^3\,|\,2 [2,2],\qquad 
{\cal WP}^3\,|\,2 [4,0].\]

\par\bigskip
\noindent
$\rm{(IV)}$\,
A super ambitwistor space (\cite[p.15]{wolf}) is defined by
\[ {\Bbb L}^{5\,|\, 2N} =
 \Bigl({\Bbb L}^5 ,\, \bigwedge^{\bullet} \bigl(({\cal O}_{{\Bbb P}^3 \times {\Bbb P}^3_*} (-1,0)\otimes {\Bbb C}^N
\oplus {\cal O}_{{\Bbb P}^3 \times {\Bbb P}^3_*} (0,-1)\otimes {\Bbb C}^N )/{\cal I}\bigr)\Bigr) ,\]
where 
\[{\cal I}=< X^{\alpha}\mu_{\alpha}-Y^{\dot{\alpha}}\lambda_{\dot{\alpha}}+2\xi_i \zeta^i >\] 
is the ideal subsheaf in 
 \[\bigwedge^{\bullet} \bigl(({\cal O}_{{\Bbb P}^3 \times {\Bbb P}^3_*} (-1,0)\otimes {\Bbb C}^N
\oplus {\cal O}_{{\Bbb P}^3 \times {\Bbb P}^3_*} (0,-1)\otimes {\Bbb C}^{N*} \bigr).
\]
The first Chern class is given by
\[ 
c_1 ({\Bbb L}^{5\,|\, 2N})=(3-N)x+(3-N)y,
\]
where $x=c_1({\cal O}_{{\Bbb L}^{5\,|\, 2N}}(1,0))$ 
and $y=c_1({\cal O}_{{\Bbb L}^{5\,|\, 2N}}(0,1))$, 
respectively (\cite{wolf}, p.20).
In the case of $N=3$, the super ambitwistor space is the super Calabi-Yau space, that is 
\[ {\Bbb L}^{5\,|\, 6} =
 \Bigl({\Bbb L}^5 ,\, \bigwedge^{\bullet} \bigl(({\cal O}_{{\Bbb P}^3 \times {\Bbb P}^3_*} (-1,0)\otimes {\Bbb C}^3 
\oplus {\cal O}_{{\Bbb P}^3 \times {\Bbb P}^3_*} (0,-1)\otimes {\Bbb C}^{3*} )/{\cal I}\bigr)\Bigr) .\]
${\cal L}^{5\,|\,6}$ is an open subset defined by 
${\cal L}^{5\,|\,6}={\Bbb L}^{5\,|\,6}\cap ({\cal P}^{3\,|\,3}\times {\cal P}^{3\,|\,3}_*)~(see \cite{ward-wells}, p.87).$ 

\end{exmp}


\section{Deformation Quantization for Super Poisson Twistor Spaces}

Let $\hat{M}$ be a supermanifold of dimension $(n\,|\,N)$. For superbivector fields $A$ and $B$ on $\hat{M}$ which are 
written locally in the form 
\[ 
A=\frac{1}{2!}\sum_{i_1,i_2}A^{i_1 i_2}\partial_{i_1}\wedge\partial_{i_2},\quad B=\frac{1}{2!}\sum_{j_1,j_2}B^{j_1 j_2}\partial_{j_1}\wedge\partial_{j_2}, 
\]
one can get a supertrivector field $[A,\,B]$ expressed locally by
\begin{align*}
 [A,B]&=\frac{1}{2!}(-1)^{|i_1 |(|j_1 |+|j_2 |+|B|)} A^{\mu i_1} \partial_{\mu} B^{j_1j_2} 
 \partial_{i_1} \wedge \partial_{j_1} \wedge \partial_{j_2}\\
&\qquad +\frac{1}{2!}(-1)^{|A |(|j_1 | +|B|)} B^{\mu j_1} \partial_{\mu} A^{i_1i_2} \partial_{i_1} 
\wedge \partial_{i_2} \wedge \partial_{j_1}. 
\end{align*}
Here $|i_k |, |j_k |, |A|, |B|$ are the Grassmann parities of the corresponding coordinates. 
The supertrivector field $[A,\,B]$ which are assigned to a pair of superbivector fields $(A,\,B)$ is called a super Schouten bracket of $A$ and $B$~
(see \cite{azcarraga-izquierdo-perelomov-bueno}). 
A super bivector $\pi$ satisfying $[\pi,\,\pi]=0$ is called a super Poisson structure. 
A super Poisson structure $\pi$ on $\hat{M}$ yields a super Poisson algebra 
in a similar way as an ordinary Poisson structure does a Poisson algebra. 

\begin{defn}
A {\rm super Poisson algebra} is a supercommutative algebra $(\mathcal{A},\,\cdot)$ endowed with a product 
\[ \{ \ ,\ \}_{{\bf Z}_2} : {\cal A}\times {\cal A}\longrightarrow {\cal A} \]
which satisfy the following conditions:
\begin{enumerate}[\quad\rm (1)]
\item $\{ f, g\}_{{\bf Z}_2} =-(-1)^{\vert f\vert \vert g \vert}\{ g,f\}_{{\bf Z}_2}$ for any homogeneous element $f,g\in {\cal A}{\rm ;}$
\item $(-1)^{\vert f \vert \vert h \vert}  \{f,\, \{ g, h \}_{{\bf Z}_2}\}_{{\bf Z}_2} +(-1)^{\vert g \vert \vert f \vert} 
    \{ g, \{ h, f \}_{{\bf Z}_2}\}_{{\bf Z}_2} +(-1)^{\vert h \vert \vert g \vert}  \{ h, \{ f, g \}_{{\bf Z}_2}\}_{{\bf Z}_2}=0{\rm ;}$
\item $\{f,\, g\cdot h \}_{{\bf Z}_2} = \{f,\,g\}_{{\bf Z}_2}\cdot h +(-1)^{\vert f \vert \vert g \vert}g\cdot \{f,\,h\}_{{\bf Z}_2}$.
\end{enumerate}
\end{defn}
The bracket $\{~,~\}_{{\bf Z}_2}$ is called a super Poisson bracket. Moreover, we call an even (resp. odd) Poisson bracket if 
$|\{ f, g\}_{{\bf Z}_2} | = |f| +|g|~(\mathrm{resp.}\, |\{ f, g\}_{{\bf Z}^2} | = |f| +|g| +1)$. 

For a super Poisson algebra $\mathcal{A}$, we denote by ${\cal A}\llbracket\hbar\rrbracket$ the space of all formal power series 
$\sum_{r=0}^{\infty} f_{r} \hbar^r$, $f_{r} \in {\cal A}$ in a parameter $\hbar$.

\begin{defn}
A super Poisson algebra $({\cal A},\cdot,\{\ ,\ \}_{{\Bbb Z}_2})$ is said to be 
{\rm deformation quantizable} if 
there exists a product 
\[ *: {\cal A}\llbracket\hbar\rrbracket\times{\cal A}\llbracket\hbar\rrbracket\longrightarrow {\cal A}\llbracket\hbar\rrbracket\]
satisfying the following conditions:
\begin{enumerate}[\quad\rm (1)]
\item $\ast$ is bilinear and $\hbar$-linear;
\item $f*(g*h)=(f*g)*h$ ~~for any $f,g,h\in {\cal A}\llbracket\hbar\rrbracket$;
\item If $f*g=f\cdot g + \sum_{r=1}^{\infty} \pi_{r} (f,g)\hbar^r$, 
then $\pi_1 (f,g)=\frac{1}{2}\{ f,g \}_{{\Bbb Z}_2}$.
\end{enumerate}
\end{defn}
The product $\ast$ is called a star product or $\ast$-product(see \cite{bffls, k, fedosov, omy, ommy1, yoshioka}).

Inspired by Definition 2.1 
for a general super manifold $\hat{M}$, 
we can define anti-chiral super manifold in the following way 
\cite{araki-takahashi-watamura, bffls, dl, fedosov, mani, manin, 
miyazaki, ommy1, ommy2, ommy3}.
\begin{defn}
Let $M=M^{4\,|\, 4N}$ be a $(4\,|\, 4N)$-dimensional super manifold, 
and with an $N$-SUSY structure equipped with a local coordinate system 
$(x^{\alpha\dot{\alpha}}\vert \theta^{i\alpha}, \theta^{\dot{\alpha}}_i ).$ 
Then $(4\,|\, 2N)$-dimensional super manifold $X$ is referred to as an 
anti-chiral super manifold, if it has anti-chiral coordinate systems 
$(x^{\alpha\dot{\alpha}}_R 
:=x^{\alpha\dot{\alpha}} 
-\theta^{i\alpha}\theta^{\dot{\alpha}}_i \vert \theta^{\dot{\alpha}}_i)$ 
with patching work. 
\end{defn}
The base space which appeared in 
the right hand side of double fibrations below studied in the present paper is   
mainly an anti-chiral complex super vector space 
${\Bbb C}^{4\,|\, 2N}$ of ${\Bbb C}^{4\,|\, 4N}$. 
We denote by $T^*_0 {\Bbb C}^{n\,|\,N}$ and $T^*_1 {\Bbb C}^{n\,|\,N}$ the even part  (${\Bbb C}^n$ bundle on ${\Bbb C}^{n\,|\,N}$) 
of the cotangent bundle $T^* {\Bbb C}^{n\,|\,N}$ and 
the odd part  (${\Bbb C}^{0\,|\,N}$ bundle on ${\Bbb C}^{n\,|\,N}$) of the cotangent bundle $T^* {\Bbb C}^{n\,|\,N}$, respectively.
They can be identified with $T^*_0 {\Bbb C}^{n\,|\,N} \simeq {\mathbb C}^{2n\,|\,N}$ and 
$ T^*_1 {\Bbb C}^{n\,|\,N} \simeq {\mathbb C}^{n\,|\,2N}$.

\begin{prop}\label{boson-fermion-commutation-relation1}
Consider a product on the even part $T^*_0 {\Bbb C}^{n\,|\,N} \simeq {\mathbb C}^{2n\,|\,N}$ of 
the cotangent bundle $T^* {\Bbb C}^{n\,|\,N}$ defined by 
\begin{equation}\label{sec3:eqn_boson-fermion-star}
f(x\,|\,\theta)*g(x\,|\,\theta) 
=
f(x\,|\,\theta) \exp\left[\frac{\hbar}{2}
\sum\frac{\overleftarrow{\partial}}{\partial X^A} E^{AB}_{\rm even} 
\frac{\overrightarrow{\partial}}{\partial X^B}\right] g(x\,|\,\theta), 
\end{equation}
where $X^A=(x^{\alpha\dot{\alpha}} \,|\, \theta_i^{\dot{\alpha}} )$ and 
\[
E^{AB}_{\rm even} =\left[\begin{array}{@{\,}cc@{\,}}
 \omega^{\alpha\dot{\alpha},\beta\dot{\beta}} & 0 \\
0 &  \omega^{i\dot{\alpha}.j\dot{\beta}}
\end{array}\right]. 
\]
\\

Then the product (\ref{sec3:eqn_boson-fermion-star}) gives deformation quantization which has the following commutation relations:  
\begin{eqnarray*}
&&[x^{\alpha \dot{\alpha}},x^{\beta \dot{\beta}}]_*=
\hbar \omega^{\alpha\dot{\alpha},\beta\dot{\beta}}, \\
&&[x^{\alpha \dot{\alpha}},\lambda_{\dot{\alpha}}]_*=[\lambda_{\dot{\alpha}},\lambda_{\dot{\beta}}]_*=0, \\
&&[x^{\alpha \dot{\alpha}},\theta_i^{\dot{\alpha}}]_*=[\lambda_{\dot{\alpha}},\theta_i^{\dot{\alpha}}]_*=0, \\
&&\{ \theta_i^{\dot{\alpha}},\theta_j^{\dot{\beta}}\}=\hbar \omega^{i\dot{\alpha},j\dot{\beta}},
\end{eqnarray*} 
where $\{~,~\}$ means a supercommutator. 
\end{prop}

\begin{prop}\label{boson-fermion-commutation-relation2}
Consider a product on the odd part $T^*_1{\Bbb C}^{n|N} \simeq {\mathbb C}^{n|2N}$ of the cotangent bundle 
$ T^*{\Bbb C}^{n|N}$ 
defined by 
\begin{equation}\label{sec3:eqn2_boson-fermion-star}
f(x\,|\,\theta)*g(x\,|\,\theta) 
=
f(x\,|\,\theta) \exp\left[\frac{\hbar}{2}
\sum\frac{\overleftarrow{\partial}}{\partial X^A} E^{AB}_{\rm odd} 
\frac{\overrightarrow{\partial}}{\partial X^B}\right] g(x\,|\,\theta), 
\end{equation}
where $X^A=(x^{\alpha\dot{\alpha}} \,|\, \theta_i^{\dot{\alpha}})$ and \\
\[
E^{AB}_{\rm odd} =\left[\begin{array}{@{\,}cc@{\,}}
 0 & \omega^{\alpha\dot{\alpha}, j \dot{\beta}}  \\
\omega^{j\dot{\beta}.\alpha \dot{\alpha}} & 0
\end{array}\right].  
\]\\
Then the product (\ref{sec3:eqn2_boson-fermion-star}) gives deformation quantization which has the following commutation relations:  
\begin{eqnarray*}
&&[x^{\alpha \dot{\alpha}},x^{\beta \dot{\beta}}]_*=0\\
&&[x^{\alpha \dot{\alpha}},\lambda_{\dot{\alpha}}]_*=[\lambda_{\dot{\alpha}},\lambda_{\dot{\beta}}]_*=0, \\
&&[x^{\alpha \dot{\alpha}},\theta_i^{\dot{\alpha}}]_*=\hbar \omega^{\alpha\dot{\alpha},j \dot{\beta}}, \\
&&[\lambda_{\dot{\alpha}},\theta_i^{\dot{\alpha}}]_*=0, \\
&&\{ \theta_i^{\dot{\alpha}},\theta_j^{\dot{\beta}}\} =0,
\end{eqnarray*} 
where $\{~,~\}$ means a supercommutator. 
\end{prop}

Using the notion of super manifold, especially ${\Bbb WP}^{n\,|\,N}$, ${\Bbb L}^{5\,|\,6}$ and their open version, 
we would like to construct and study star products with respect to them in the sense of the previous paragraph. 
One may see that super twistor double fibrations play crucial roles.

\begin{thm}\label{boson-fermion-commutation-relation3}
Let $({\mathbb C}^{4\,|\,8}\times {\mathbb P}^1, {\cal O}_{{\mathbb C}^{4\,|\,8}\times {\mathbb P}^1}, \pi)$ be one of super Poisson algebra 
on the super Calabi-Yau twistor space ${\mathbb C}^{4\,|\,8}\times {\mathbb P}^1$  
defined in the following way{\rm :} 
consider a diagram 
\[
\begin{array}{@{\,}ccccc@{\,}}
{\cal P}^{3\,|\,4} & \longleftarrow & {\mathbb C}^{4\,|\,8}\times {\mathbb P}^1 
& \longrightarrow & {\mathbb C}^{4\,|\,8} \\ 

[x_R^{\alpha \dot{\alpha}}  
\lambda_{\dot{\alpha}} , \lambda_{\dot{\alpha}} \,|\, \theta_i^{\dot{ \alpha}} \lambda_{\dot{\alpha}}] 
& \longleftarrow & 
(x_R^{\alpha\dot{\alpha}} ,\lambda_{\dot{\alpha}} \,|\, \theta_i^{\dot{\alpha}}) & 
\longrightarrow & 
(x_R^{\alpha\dot{\alpha}} \,|\, \theta_i^{\dot{\alpha}}).
\end{array}
\]
and a bivector 
\[ \pi = \lambda_{\dot{1}}\lambda_{\dot{2}}\frac{\partial}{\partial z^1}\wedge
\frac{\partial}{\partial z^2} 
+\frac{1}{2}\,\bigl\{(\lambda_{\dot{1}})^2 +( \lambda_{\dot{2}})^2\bigr\}
\sum_{i=1}^4 \frac{\partial}{\partial \xi_i}\wedge
\frac{\partial}{\partial \xi_i} ,\]
where $z^{\alpha}=x_R^{\alpha \dot{\alpha}}\lambda_{\dot{\alpha}},  
\xi_i= \theta_i^{\dot{ \alpha}} \lambda_{\dot{\alpha}}.$ 

Then $({\cal O}_{{\mathbb C}^{4\,|\,8}\times {\mathbb P}^1}, \pi)$ can be deformation quantizable
and the non-commutative product $*$ 
is explicitly written down in the following. 
\begin{equation}\label{boson-fermion-star}
f(z\,|\,\xi)*g(z\,|\,\xi) 
=
f(z\,|\,\xi) \exp\left[\frac{\hbar}{2}
\sum\frac{\overleftarrow{\partial}}{\partial X^A} \pi^{AB} 
\frac{\overrightarrow{\partial}}{\partial X^B}\right] g(z\,|\,\xi), 
\end{equation}
where $X^A =(z^{\alpha},\lambda_{\dot{\alpha}}\, |\, \xi_i )$ or 
$X^A =(X^{\alpha}, \lambda_{\dot{\alpha}}, Y^{\dot{\alpha}}, \mu_{\alpha}\, |\, \xi_i, \zeta^i )$ 
$\in {\mathbb C}^{4\,|\,8}\times {\mathbb P}^1$.
$\pi^{AB}$ is one of the super even Poisson bivectors 
on the super Calabi-Yau twistor family $ {\mathbb C}^{4\,|\,8}\times {\mathbb P}^1$. 
\end{thm}

\begin{pf}

First, we remark that the bivector satisfies $[\pi,\pi]=0$ 
since the coefficients of each term do not have variables 
which are differentiated by $\pi$.  
Therefore $\pi$ is a super even Poisson bivector. 
Next we would like to study quantization of $( {\mathbb C}^{4\,|\,8}\times {\mathbb P}^1,\pi)$.  
When we put $z^{\alpha}=x_R^{\alpha\dot{\alpha}}\lambda_{\dot{\alpha}}$, 
$\xi_i =\theta_i^{\dot{\alpha}}\lambda_{\dot{\alpha}}$,
the local coordinates on ${\cal P}^{3\,|\,4}$ satisfy the following 
commutation relations{\rm :}
\begin{eqnarray*}
&&[z^1,z^2]_*=
2\hbar \lambda_{\dot{1}}\lambda_{\dot{2}}, \\
&&\{ \xi_i,\xi_i \}_* =\hbar ((\lambda_{\dot{1}})^2 +( \lambda_{\dot{2}})^2), \\&& (0 \ \ \rm{o.w.}). 
\end{eqnarray*}

We first prove for ${\cal P}^{3|4}= {\cal O}_{{\Bbb P}^1}(1) \oplus {\cal O}_{{\Bbb P}^1}(1) 
\oplus \bigoplus_{j=1}^4 \Pi {\cal O}_{{\Bbb P}^1}(1)$.
Consider the following double fibration (cf. \cite{taniguchi}).

\begin{equation} 
\begin{array}{ccccc}
{\cal P}^{3\,|\,N} & \longleftarrow & {\mathbb C}^{4\,|\,2N}\times {\mathbb P}^1 
& \longrightarrow & {\mathbb C}^{4\,|\,2N} \\ 
{[} x_R^{\alpha \dot{\alpha}}  
\lambda_{\dot{\alpha}}, \lambda_{\dot{\alpha}} \,|\, \theta_i^{ \dot{\alpha}} \lambda_{\dot{\alpha}} ] 
& \longleftarrow & 
(x_R^{\alpha\dot{\alpha}} ,\lambda_{\dot{\alpha}} \,|\, \theta_i^{\dot{\alpha}}) & 
\longrightarrow & 
(x_R^{\alpha\dot{\alpha}} \,|\, \theta_i^{\dot{\alpha}}).
\end{array}
\end{equation}
If we put $z^{\alpha}:=x^{\alpha\dot{\alpha}}\lambda_{\dot{\alpha}}$, $\xi_i:=\theta_i^{\dot{\alpha}}\lambda_{\dot{\alpha}}$, 
then we see 
\begin{eqnarray*}
[z^{\alpha},z^{\beta}]_\ast
&=&[x^{\alpha\dot{\alpha}} \lambda_{\dot{\alpha}},  
   x^{\beta\dot{\beta}} \lambda_{\dot{\beta}}]_\ast \\
&=&[x^{\alpha\dot{\alpha}}, x^{\beta\dot{\beta}}]_\ast
   \lambda_{\dot{\alpha}} \lambda_{\dot{\beta}} \\
&=&\hbar D^{\alpha\dot{\alpha},\beta\dot{\beta}}
   \lambda_{\dot{\alpha}} \lambda_{\dot{\beta}}  
\end{eqnarray*}
and
\begin{eqnarray*}
\{ \xi_i,\xi_j \}_* &=& \{ \theta_i^{\dot{\alpha}}\lambda_{\dot{\alpha}}, \theta_j^{\dot{\beta}}\lambda_{\dot{\beta}} \}_* \\
&=& \{ \theta_i^{\dot{\alpha}}, \theta_j^{\dot{\beta}}\}_* \lambda_{\dot{\alpha}}\lambda_{\dot{\beta}} \\
&=& \hbar C^{i\dot{\alpha}, j\dot{\beta}}\lambda_{\dot{\alpha}}\lambda_{\dot{\beta}},
\end{eqnarray*}
where we used 
$[x^{\alpha\dot{\alpha}}, x^{\beta\dot{\beta}}]_\ast
=\hbar D^{\alpha\dot{\alpha},\beta\dot{\beta}}$
and 
 $\{\theta_i^{\dot{\alpha}},\theta_j^{\dot{\beta}}\}_*=\hbar 
C^{i\dot{\alpha},j\dot{\beta}}\lambda_{\dot{\alpha}}\lambda_{\dot{\beta}}$.

This indicates the star-product is globally defined (cf. the proof of 
Theorem~\ref{boson-fermion-commutation-relation3}). 
We would like to show the equality above makes sense by patching together. 
Let ${\cal P}^{3\,|\,4} \ni (z^{\alpha}, \lambda_{\dot{\alpha}} \,|\,\xi^i)$ 
be the homogeneous coordinate. Using this coordinate, we introduce 
the coordinate neighborhood in the following way.

\begin{equation}
\omega_+^{\alpha}:=\frac{z^{\alpha} }{\lambda_{\dot{1}}},\quad
\lambda_+:=\frac{\lambda_{\dot{2}}}{\lambda_{\dot{1}}},\quad \xi_+^i:=\frac{\xi_i}{\lambda_{\dot{1}}} \quad 
{\rm on~}{\cal U}_+ ,  
\end{equation}
\begin{equation}
\omega_-^{\alpha}:=\frac{z^{\alpha} }{\lambda_{\dot{2}}},\quad
\lambda_-:=\frac{\lambda_{\dot{1}}}{\lambda_{\dot{2}}},\quad \xi_-^i:=\frac{\xi_i}{\lambda_{\dot{2}}}\quad 
{\rm on~}{\cal U}_-.   
\end{equation}
The transition functions are given in the following way. 
\begin{equation}
\omega_-^{\alpha}=\frac{1}{\lambda_+}\omega_+^{\alpha},
\quad \lambda_-=\frac{1}{\lambda_+},\quad \xi_-^i=\frac{1}{\lambda_+}\xi_+^i,\quad 
{\rm on~} {\cal P}^{3\,|\,4}
={\cal U}_+\cup{\cal U}_-. 
\end{equation}
Under the coordinate above, we have 
\begin{equation}\label{maruichi}
\{\xi_+^i,\xi_+^j\}_* = \hbar\bigr(C^{i\dot{1},j\dot{1}} + 2C^{i\dot{1},j\dot{2}}\lambda_+ + C^{i\dot{2},j\dot{2}}\lambda_+\lambda_+\bigr) \quad{\rm on~}{\cal U}_+,
\end{equation}
\begin{equation}\label{maruni}
\{\xi_-^i,\xi_-^j\}_* = \hbar\bigr(C^{i\dot{1},j\dot{1}}\lambda_-\lambda_- + 2C^{i\dot{1},j\dot{2}}\lambda_- + C^{i\dot{2},j\dot{2}}
\bigr) \quad{\rm on~}{\cal U}_-.
\end{equation}
Thus we obtain 
\begin{equation}
\{\xi_-^i,\xi_-^j\}_*=\left( \frac{1}{\lambda_+} \right)^2\, \{\xi_+^i,\xi_+^j\}_*. 
\end{equation}

\end{pf}

The arguments similar 
to this can be applied for patching works of super Poisson Calabi-Yau,  
${\cal WP}^{2\,|\,3} [1,3]$, ${\cal WP}^{2\,|\,3}
 [2,2]$, ${\cal WP}^{2\,|\,3} [4,0]$, ${\cal L}^{5\,|\,6}.$

\section{Other examples} 
In the present section we give other examples. 

\begin{prop}\label{i}
A complex super projective space of dimension $(n\vert N)$ 
is defined by 
\[
 {\Bbb P}^{n\vert N}=({\Bbb P}^n, \bigwedge^{\cdot}
({\Bbb C}^N \otimes_{\Bbb C} {\cal O}_{{\Bbb P}^n}(-1))).
\]
We denote by $ {\cal O}_{{\Bbb P}^{n\vert N}}$ the structure sheaf  
$\bigwedge^{\cdot} ({\Bbb C}^N \otimes_{\Bbb C}{\cal O}_{{\Bbb P}^n}(-1))$
of ${\Bbb P}^{n\vert N}$. 
A local section of ${\cal O}_{{\Bbb P}^{n\vert N}}$ is represented 
as in (\ref{local-represent}) via a degree $(-k)$ homogeneous element 
$f_{i_1 i_2\ldots i_k}(z)$. 
For example, if $(n|N)=(3|2)$, we see that 
${\cal O}_{{\Bbb P}^{3\vert 2}}$ is decomposed in the following way: 
\[ {\cal O}_{{\Bbb P}^{3\vert 2}}\cong {\cal O}_{{\Bbb P}^3} \oplus 
\Pi {\cal O}_{{\Bbb P}^3}(-1)\oplus \Pi {\cal O}_{{\Bbb P}^3}(-1)\oplus 
{\cal O}_{{\Bbb P}^3}(-2), \]
where $\Pi$ denotes the parity change functor.
We define the line sheaf of degree $d$ that 
${\cal O}_{{\Bbb P}^{n\vert N}} (d) = 
{\cal O}_{{\Bbb P}^n}(d)\otimes {\cal O}_{{\Bbb P}^{n\vert N}}$, 
where 
we denote the sheaf of germs 
of locally defined, holomorphic, homogeneous degree $d$ 
by 
${\cal O}_{\Bbb P}^n$
The first Chern class is given by
\[ 
c_1 ({\Bbb P}^{n \vert N})=
c_1 ({\cal O}_{{\Bbb P}^{n \vert N}} \otimes 
{\Bbb C}^{n+1}) 
- c_1 ({\cal O}_{{\Bbb P}^{n \vert N}}\otimes 
{\Bbb C}^N) = (n+1-N )x, 
\]
where $x=c_1 ({\cal O}_{{\Bbb P}^{n\vert N}}(1))$.
Hence we may conclude that super twistor space 
${\Bbb P}^{n\vert N}$ 
becomes a Calabi-Yau supermanifold if and only if $N=4$.

Deformation quantization for this object is 
constructed by the following diagram :
\[
\begin{array}{ccccc}
{\cal WP}^{3|2} [1,3]& \longleftarrow & {\mathbb C}^{4|6}\times {\mathbb P}^1 
& \longrightarrow & {\mathbb C}^{4|6} \\ 

{[} x_R^{\alpha \dot{\alpha}}  
\lambda_{\dot{\alpha}} , \lambda_{\dot{\alpha}} | \theta_1^{\dot{ \alpha}_1} \lambda_{\dot{\alpha}_1}, \theta_2^{\dot{ \alpha}_1 \dot{ \alpha}_2\dot{ \alpha}_3 } 
\lambda_{\dot{\alpha}_1}\lambda_{\dot{\alpha}_2} \lambda_{\dot{\alpha}_3}  ] 
& \longleftarrow & 
(x_R^{\alpha\dot{\alpha}} ,\lambda_{\dot{\alpha}} | \theta_1^{\dot{\alpha}_1} ,\theta_2^{\dot{\alpha}_1\dot{\alpha}_2\dot{\alpha}_3  }) & 
\longrightarrow & 
(x_R^{\alpha\dot{\alpha}}  | \theta_1^{\dot{\alpha}_1} ,\theta_2^{\dot{\alpha}_1\dot{\alpha}_2\dot{\alpha}_3  }) .
\end{array}
\]

We define a super even bivector as follows.
\[ \pi = \lambda_{\dot{1}}\lambda_{\dot{2}}\frac{\partial}{\partial z^1}\wedge\frac{\partial}{\partial z^2} +\frac{1}{2}((\lambda_{\dot{1}})^2 +( \lambda_{\dot{2}})^2)
 \frac{\partial}{\partial \xi_1}\vee\frac{\partial}{\partial \xi_1} +\frac{1}{2}((\lambda_{\dot{1}})^2 +( \lambda_{\dot{2}})^2)^3
 \frac{\partial}{\partial \xi_2}\vee\frac{\partial}{\partial \xi_2} . \]
The bivector $\pi$ satisfies $[\pi,\pi]=0$ 
with respect to super Schouten bracket 
because coefficients of each term do not have variables 
which are differentiated by $\pi$.  
Then $\pi$ defines a super even Poisson bivector.

When we put $z^{\alpha}=x_R^{\alpha\dot{\alpha}}\lambda_{\dot{\alpha}}$, 
$\xi_1 =\theta_1^{\dot{\alpha}_1}\lambda_{\dot{\alpha}_1}$,
$\xi_2 =\theta_2^{\dot{\alpha}_1\dot{\alpha}_2\dot{\alpha}_3}\lambda_{\dot{\alpha}_1}\lambda_{\dot{\alpha}_2}\lambda_{\dot{\alpha}_3}$,
those commutation relations is the following.
\begin{eqnarray*}
&&[z^1,z^2]_*=
2\hbar \lambda_{\dot{1}}\lambda_{\dot{2}}, \\
&&\{ \xi_1,\xi_1 \}_* =\hbar ((\lambda_{\dot{1}})^2 +( \lambda_{\dot{2}})^2),\\
&&\{ \xi_2,\xi_2 \}_*  =\hbar ((\lambda_{\dot{1}})^2 +( \lambda_{\dot{2}})^2)^3,\\
&& (0 \ \ \rm{o.w.}). 
\end{eqnarray*} 

\end{prop}

\begin{prop}\label{iii}
A weighted super projective space (\cite{wolf}, p.40) 
is defined by
\[
{\Bbb W}{\Bbb P}^{n\vert N} [k_1,k_2,\dots , k_{n+1}\vert l_1,l_2,\dots ,l_N ]
=  ({\Bbb W} {\Bbb P}^n [k_1,k_2,\dots ,k_{n+1}] , 
{\cal O}_{{\Bbb W}{\Bbb P}^{n\vert N}} ), 
\]
where 
\[  {\cal O}_{{\Bbb W}{\Bbb P}^{n\vert N} }
= \bigwedge^{\cdot} ({\cal O}_{{\Bbb W}{\Bbb P}^n} (-l_1)\oplus \dots 
\oplus {\cal O}_{{\Bbb W}{\Bbb P}^n} (-l_N )) .\]
Note that with the definitions above, we have   
\[ {\Bbb W}{\Bbb P}^{n\vert N} 
[\overbrace{1,1, \dots , 1}^{n+1} \vert \overbrace{1,1,\dots ,1}^N ] = {\Bbb P}^{n\vert N} .\]
The first Chern class is given by
\[ c_1 ({\Bbb W}{\Bbb P}^{n\vert N})=(\sum_{i=1}^{n+1}k_i - \sum_{j=1}^{N} l_j )x, \]
$x=c_1 ({\cal O}_{{\Bbb W}{\Bbb P}^{n\vert N}}(1))$.
Hence, for appropriate numbers $k_i$ and $l_j$, the weighted projective super space ${\Bbb W}{\Bbb P}^{n\vert N}$
becomes a Calabi-Yau supermanifold.\\

Let us now consider an open subset of ${\Bbb W}{\Bbb P}^{3\vert N} [k_1,k_2,1,1 \vert l_1,l_2,\dots ,l_N ]$ defined by
\begin{equation}
\label{open-wp} 
\begin{array}{ll} 
& {\cal WP}^{3|N} [k_1,k_2,1,1 \vert l_1,l_2,\dots ,l_N] \\ 
= & {\Bbb W}{\Bbb P}^{3\vert N} [k_1,k_2, 1, 1 \vert l_1,l_2,\dots , l_N ] 
\setminus 
{\Bbb W}{\Bbb P}^{1\vert N} [k_1,k_2,1, 1 \vert  l_1,l_2,\dots , l_N ]. 
\end{array}
\end{equation}
The space defined as above can be identified with  the holomorphic fiber bundle
 \[  {\cal O}_{{\Bbb P}^1}(k_1) \oplus {\cal O}_{{\Bbb P}^1}(k_2) \oplus \bigoplus_{j=1}^N \Pi {\cal O}_{{\Bbb P}^1}(l_j) \longrightarrow {\Bbb P}^1, \]
and as such it can be covered by two patches (cf. \cite{wolf}).

In the case of $k_1=k_2=l_1=l_2=\dots =l_N=1$,
 \[ {\cal P}^{3|N}= {\cal O}_{{\Bbb P}^1}(1) \oplus {\cal O}_{{\Bbb P}^1}(1) \oplus \bigoplus_{j=1}^N \Pi {\cal O}_{{\Bbb P}^1}(1), \quad 
{\cal P}^{3|N} \cong {\Bbb P}^{3|N}\setminus {\Bbb P}^{1|N}. \] 

In this case, Calabi-Yau condition is of the $k_1+k_2-(l_1+\dots l_N)+2=0$.\\
In particular we consider the case of $N=2$, that is,  

\[{\cal WP}^{3|2} [p,q]={\Bbb W}{\Bbb P}^{3\vert 2} [1, 1, 1, 1 \vert p,q ]\setminus {\Bbb W}{\Bbb P}^{1\vert 2} [1, 1 \vert p,q ]. \]
This space can be identified with  the holomorphic fiber bundle
 \[  {\cal O}_{{\Bbb P}^1}(1) \oplus {\cal O}_{{\Bbb P}^1}(1) \oplus 
\Pi {\cal O}_{{\Bbb P}^1}(p)\oplus \Pi {\cal O}_{{\Bbb P}^1}(q)\longrightarrow {\Bbb P}^1. \]
For the particular combination $p+q=4$, 
it becomes a Calabi-Yau supermanifold.\\
There are three types in the case of Calabi-Yau supermanifold. It is that
\[ {\cal WP}^{2|3} [1,3],\qquad  
{\cal WP}^{2|3} [2,2],\qquad 
{\cal WP}^{2|3} [4,0].\]

A super ambitwistor space (\cite{wolf}, p.15) is defined by
\[ {\Bbb L}^{5\vert 2N} =
 ({\Bbb L}^5 , \bigwedge^{\cdot} (({\cal O}_{{\Bbb P}^3 \times {\Bbb P}^3_*} (-1,0)\otimes {\Bbb C}^N
\oplus {\cal O}_{{\Bbb P}^3 \times {\Bbb P}^3_*} (0,-1)\otimes {\Bbb C}^N )/{\cal I})) ,\]
where 
\[{\cal I}=< X^{\alpha}\mu_{\alpha}-Y^{\dot{\alpha}}\lambda_{\dot{\alpha}}+2\xi_i \zeta^i >\] 
is the ideal subsheaf in 
 \[\wedge^{\cdot} (({\cal O}_{{\Bbb P}^3 \times {\Bbb P}^3_*} (-1,0)\otimes {\Bbb C}^N
\oplus {\cal O}_{{\Bbb P}^3 \times {\Bbb P}^3_*} (0,-1)\otimes {\Bbb C}^{N*} ).
\]
The first Chern class is given by
\[ c_1 ({\Bbb L}^{5\vert 2N})=(3-N)x+(3-N)y,\]
where $x=c_1({\cal O}_{{\Bbb L}^{5\vert 2N}}(1,0))$ 
and $y=c_1({\cal O}_{{\Bbb L}^{5\vert 2N}}(0,1))$, 
respectively (\cite{wolf}, p.20).
In the case of $N=3$, the super ambitwistor space is the super Calabi-Yau space, that is
\[ {\Bbb L}^{5\vert 6} =
 ({\Bbb L}^5 , \bigwedge^{\cdot} (({\cal O}_{{\Bbb P}^3 \times {\Bbb P}^3_*} (-1,0)\otimes {\Bbb C}^3 
\oplus {\cal O}_{{\Bbb P}^3 \times {\Bbb P}^3_*} (0,-1)\otimes {\Bbb C}^{3*} )/{\cal I})) .\]

${\cal L}^{5|6}$ is an open subset defined by 
${\cal L}^{5|6}={\Bbb L}^{5|6}\cap ({\cal P}^{3|3}\times {\cal P}^{3|3}_*) \ [24 p.87].$ 

Star product for this object is constructed by the following diagram :
\[
\begin{array}{ccccc}
{\cal WP}^{3|2}[2,2] & \longleftarrow & {\mathbb C}^{4|6}\times {\mathbb P}^1 
& \longrightarrow & {\mathbb C}^{4|6} \\ 

{[} x_R^{\alpha \dot{\alpha}} , 
\lambda_{\dot{\alpha}}  | \theta_1^{\dot{ \alpha}_1\dot{ \alpha}_2} \lambda_{\dot{\alpha}_1} \lambda_{\dot{\alpha}_2}, 
\theta_2^{\dot{ \alpha}_1\dot{ \alpha}_2} \lambda_{\dot{\alpha}_1} \lambda_{\dot{\alpha}_2} ] 
& \longleftarrow & 
(x_R^{\alpha\dot{\alpha}}\lambda_{\dot{\alpha}} ,\lambda_{\dot{\alpha}} | \theta_1^{\dot{\alpha}_1\dot{\alpha}_2},  \theta_2^{\dot{\alpha}_1\dot{\alpha}_2} ) & 
\longrightarrow & 
(x_R^{\alpha\dot{\alpha}} | \theta_1^{\dot{\alpha}_1\dot{\alpha}_2}, \theta_2^{\dot{\alpha}_1\dot{\alpha}_2} ).
\end{array}
\]

We introduce a bivector as follows:
\[ \pi = \lambda_{\dot{1}}\lambda_{\dot{2}}\frac{\partial}{\partial z^1}\wedge\frac{\partial}{\partial z^2} 
+\frac{1}{2}((\lambda_{\dot{1}})^2 +( \lambda_{\dot{2}})^2)^2
(\frac{\partial}{\partial \xi_1}\vee\frac{\partial}{\partial \xi_1} +\frac{\partial}{\partial \xi_2}\vee\frac{\partial}{\partial \xi_2} ) . \]

When we put $z^{\alpha}=x_R^{\alpha\dot{\alpha}}\lambda_{\dot{\alpha}}$, 
$\xi_1 =\theta_1^{\dot{\alpha}_1\dot{\alpha}_2} \lambda_{\dot{\alpha}_1}\lambda_{\dot{\alpha}_2}$,
$\xi_2 =\theta_2^{\dot{\alpha}_1\dot{\alpha}_2}\lambda_{\dot{\alpha}_1}\lambda_{\dot{\alpha}_2}$,
those commutation relations are the followings.
\begin{eqnarray*}
&&[z^1,z^2]_*=
2\hbar \lambda_{\dot{1}}\lambda_{\dot{2}}, \\
&&\{ \xi_1,\xi_1 \}_* =\{\xi_2,\xi_2 \}_* = \hbar ((\lambda_{\dot{1}})^2 +( \lambda_{\dot{2}})^2)^2,\\
&& (0 \ \ \rm{o.w.}).
\end{eqnarray*}

\end{prop}

\begin{prop}\label{iv}
For this object, we use the following diagram :
\[
\begin{array}{ccccc}
{\cal WP}^{3|2} [4,0]& \longleftarrow & {\mathbb C}^{4|6}\times {\mathbb P}^1 
& \longrightarrow & {\mathbb C}^{4|6} \\ 
{[} x_R^{\alpha \dot{\alpha}}  
\lambda_{\dot{\alpha}} , \lambda_{\dot{\alpha}} | \theta_1^{\dot{ \alpha}_1 \dot{ \alpha}_2\dot{ \alpha}_3\dot{ \alpha}_4    }
 \lambda_{\dot{\alpha}_1}  \lambda_{\dot{\alpha}_2} \lambda_{\dot{\alpha}_3} \lambda_{\dot{\alpha}_4}, \theta_2   ] 
& \longleftarrow & 
(x_R^{\alpha\dot{\alpha}} ,\lambda_{\dot{\alpha}} | \theta_1^{\dot{\alpha}_1 \dot{\alpha}_2\dot{\alpha}_3\dot{\alpha}_4 } ,\theta_2   ) & 
\longrightarrow & 
(x_R^{\alpha\dot{\alpha}} | \theta_1^{\dot{\alpha}_1 \dot{\alpha}_2\dot{\alpha}_3\dot{\alpha}_4 } ,\theta_2       ).
\end{array}
\]

We introduce an even Poisson bivector as follows.
\[ \pi = \lambda_{\dot{1}}\lambda_{\dot{2}}\frac{\partial}{\partial z^1}\wedge\frac{\partial}{\partial z^2} +\frac{1}{2}((\lambda_{\dot{1}})^2 +( \lambda_{\dot{2}})^2)^4
 \frac{\partial}{\partial \xi_1}\vee\frac{\partial}{\partial \xi_1} +\frac{1}{2}
 \frac{\partial}{\partial \xi_2}\vee\frac{\partial}{\partial \xi_2} . \]
Obviously, the bivector $\pi$ satisfies $[\pi,\pi]=0$ 
with respect to super Schouten bracket. 

When we put 
$z^{\alpha}=x_R^{\alpha\dot{\alpha}}\lambda_{\dot{\alpha}}$, 
$\xi_1 =\theta_1^{\dot{\alpha}_1\dot{\alpha}_2 \dot{\alpha}_3 \dot{\alpha}_4} \lambda_{\dot{\alpha}_1}\lambda_{\dot{\alpha}_2}\lambda_{\dot{\alpha}_3}\lambda_{\dot{\alpha}_4}$,
$\xi_2 =\theta_2$,
the $*$-product satisfies the following commutation relations:
\begin{eqnarray*}
&&[z^1,z^2]_*=
2\hbar \lambda_{\dot{1}}\lambda_{\dot{2}}, \\
&&\{ \xi_1,\xi_1 \}_* =
\hbar ((\lambda_{\dot{1}})^2 +( \lambda_{\dot{2}})^2)^4,\\
&&\{ \xi_2,\xi_2 \}_*  =\hbar, \\
&& (0 \ \ \rm{o.w.}). 
\end{eqnarray*} 

\end{prop}

\begin{prop}\label{v}
The fifth example is constructed using the following diagram :. 
\[{\Bbb C}^{4|24} \ni (x^{\mu} | \Theta^A, \Theta^{\dot{A}}) \mapsto 
(x^{\mu}_R := x^{\mu} - a_{A\dot{B}}^{\mu}\Theta^A\Theta^{\dot{B}} | \Theta^{\dot{A}}) \in {\Bbb C}^{4|12}. \]

If we put $x^{\alpha\dot{\alpha}} = x_R^{\mu}$, $(\theta_i^{\dot{\alpha}}, \eta^{i\alpha} ) =\Theta^{\dot{A}}$,
$\dot{A} =\dot{1},\dots ,\dot{12}$, $i=1,2,3$,
then the anti-chiral SUSY structure can be defined by
\[ \hat{Q}_{\dot{A}} =\frac{\partial}{\partial\Theta^{\dot{A}}}, 
\qquad \hat{Q}_A = \frac{\partial}{\partial\Theta^A} +2a^{\mu}_{A\dot{B}}\Theta^{\dot{B}}\frac{\partial}{\partial x^{\mu}} .\]
Hence we have
\[
\begin{array}{ccccc}
{\cal L}^{5|6} & \longleftarrow & {\mathbb C}^{4|12}\times {\mathbb P}^1\times{\mathbb P}_*^1 
& \longrightarrow & {\mathbb C}^{4|12} \\ 

{[} X^{\alpha},\lambda_{\dot{\alpha}},Y^{\dot{\alpha}}, \mu_{\alpha} | \xi_i, \zeta^i   ] 
& \longleftarrow & 
(x^{\alpha\dot{\alpha}} ,\lambda_{\dot{\alpha}} , \mu_{\alpha} | \theta_i^{\dot{\alpha}}, \eta^{i\alpha} ) & 
\longrightarrow & 
(x^{\alpha\dot{\alpha}} | \theta_i^{\dot{\alpha}}, \eta^{i\alpha} ).
\end{array}
\]

We define a super even bivector as follows.
\[ 
\begin{array}{lll}
\pi& = &\lambda_{\dot{1}}\lambda_{\dot{2}}\frac{\partial}{\partial X^1}
\wedge\frac{\partial}{\partial X^2} \\ 
&+&\frac{1}{2}\lambda_{\dot{2}}\mu_2 \frac{\partial}{\partial X^1} 
\wedge \frac{\partial}{\partial Y^{\dot{1}}} \\
 &+& \frac{1}{2}\lambda_{\dot{1}}\mu_2 \frac{\partial}{\partial X^1} 
\wedge \frac{\partial}{\partial Y^{\dot{2}}}  
 -\frac{1}{2}\lambda_{\dot{2}}\mu_1 \frac{\partial}{\partial X^2} 
\wedge\frac{\partial}{\partial Y^{\dot{1}}}
 -\frac{1}{2}\lambda_{\dot{1}}\mu_1 \frac{\partial}{\partial X^2}
\wedge\frac{\partial}{\partial Y^{\dot{2}}} \\  
 &+&\frac{1}{2}((\lambda_{\dot{1}})^2 +( \lambda_{\dot{2}})^2)
\sum_{i=1}^3 \frac{\partial}{\partial \xi_i}\vee\frac{\partial}{\partial \xi_i} +\frac{1}{2}((\mu_1 )^2 +( \mu_2 )^2)
\sum_{i=1}^3 \frac{\partial}{\partial \zeta^i}\vee\frac{\partial}{\partial \zeta^i} . 
\end{array}
\]

When we put\\
$[ X^{\alpha}, \lambda_{\dot{\alpha}}, Y^{\dot{\alpha}},  \mu_{\alpha}  |  \xi_i,  \zeta^i ]  =          
 [ (x^{\alpha \dot{\alpha}} -\eta^{i\alpha}\theta_i^{\dot{\alpha}})\lambda_{\dot{\alpha}}, \lambda_{\dot{\alpha}}, 
 (x^{\alpha \dot{\alpha}} +\theta_i^{\dot{\alpha}}\eta^{i\alpha})\mu_{\alpha},\mu_{\alpha} 
 | \theta_i^{\dot{ \alpha}} \lambda_{\dot{\alpha}}, \eta^{i\alpha} \mu_{\alpha} ] $\\
then the $*$-product satisfies the following commutation relations.
\begin{eqnarray*}
&&[X^1,Y^{\dot{1}} ]_*=
\hbar \lambda_{\dot{2}}\mu_2, \\
&&[X^1,Y^{\dot{2}} ]_*=
\hbar \lambda_{\dot{1}}\mu_2, \\
&&[X^2,Y^{\dot{1}} ]_*=
-\hbar \lambda_{\dot{2}}\mu_1, \\
&&[X^2,Y^{\dot{2}} ]_*=
-\hbar \lambda_{\dot{1}}\mu_1, \\
&&\{ \xi_i,\xi_i \}_* =\hbar ((\lambda_{\dot{1}})^2 +( \lambda_{\dot{2}})^2) \ (i=1,2,3),\\
&&\{ \zeta_i,\zeta_i \}_* =\hbar ((\mu_1)^2 +( \mu_2)^2) \ (i=1,2,3),\\
&& [X^1, X^2 ]_* =2\hbar \lambda_{\dot{1}}\lambda_{\dot{2}},\\
&& [Y^{\dot{1}}, Y^{\dot{2}} ]_* =0, \\
&& (0 \ \ \rm{o.w.}). 
\end{eqnarray*} 

\end{prop}

\section{Glueing}

Next we would like to show the above argument can be extended to 
${\mathbb P}^{3|N}$. 
Replacing $(z^{1},z^{2},\lambda_{\dot{1}},\lambda_{\dot{2}} |\xi_i)$ 
by $(z_1,z_2,z_3,z_4|\xi_i) (i=1\sim N)$, we see that 
\begin{equation}\label{maruzero}
\{\xi_i,\xi_j\}_*=\hbar\bigr(C^{i\dot{1},j\dot{1}}z_3z_3+2C^{i\dot{1},j\dot{2}}z_3z_4+C^{i\dot{2},j\dot{2}}z_4z_4
\bigr). 
\end{equation}
Take an inhomogeneous coordinate system in the following way: 
\begin{eqnarray}
&& \left( \frac{z_2}{z_1},\frac{z_3}{z_1},\frac{z_4}{z_1}\Bigr|
\frac{\xi_1}{z_1},\ldots,\frac{\xi_N}{z_1}  \right)\mbox{ on }U_1,\\
&&\nonumber\quad\quad\quad\quad\quad\vdots\\
&& \left( \frac{z_1}{z_4},\frac{z_2}{z_4},\frac{z_3}{z_4}\Bigr|
\frac{\xi_1}{z_4},\ldots,\frac{\xi_N}{z_4}  \right)\mbox{ on }U_4. 
\end{eqnarray}
Set $z_l^k:=\frac{z_k}{z_l},~\xi_l^i:=\frac{\xi_i}{z_l}$.  
Then 
\begin{eqnarray}
&&z_k^m=\frac{1}{z_l^k}z_l^m,~~z_k^l=\frac{1}{z_l^k},~~
\xi_k^i=\frac{1}{z_l^k}\xi_l^i \mbox{ on } U_k\cap U_l(k\not=m,~l\not=m), \\
&&{\mathbb P}^{3|N}=U_1\cup U_2\cup U_3\cup U_4. 
\end{eqnarray}
Using this coordinate system and transition functions, we see that 
\begin{eqnarray}
&&\label{maruichi-}
\{\xi_k^i,\xi_k^j\}_*=\hbar\bigr(C^{i\dot{1},j\dot{1}}\frac{z_3}{z_k}\frac{z_3}{z_k}
+2C^{i\dot{1},j\dot{2}}\frac{z_3}{z_k}\frac{z_4}{z_k}+C^{i\dot{2},j\dot{2}}
\frac{z_4}{z_k}\frac{z_4}{z_k}\bigr),\\
&&\label{maruni-}
\{\xi_l^i,\xi_l^j\}_*=\hbar\bigr(C^{i\dot{1},j\dot{1}}\frac{z_3}{z_l}\frac{z_3}{z_l}
+2C^{i\dot{1},j\dot{2}}\frac{z_3}{z_l}\frac{z_4}{z_l}+C^{i\dot{2},j\dot{2}}
\frac{z_4}{z_l}\frac{z_4}{z_l}\bigr).
\end{eqnarray}
Hence we obtain the following identity. 
$$(\ref{maruni-})=\left(\frac{1}{z_l^k}\right)^2(\ref{maruichi-}). $$
This indicates well-definedness of the product $f*g$ 
for sections 
$f,g\in {\cal O}_{{\mathbb P}^{3|N}}=\wedge({\mathbb C}^N\otimes_{\mathbb C} 
{\cal O}_{{\mathbb P}^3}(-1))$ which represent germes. 

Taking a real slice we can show the same formula for case of  
$R={\mathbb R}$. \\

\end{document}